\documentclass[12pt]{article}
\begin{document}

\title{{\bf Normalized Observational Probabilities from Unnormalizable Quantum States or Phase-Space Distributions}
\thanks{Alberta-Thy-5-12, arXiv:yymm.nnnn [hep-th]}}

\author{
Don N. Page
\thanks{Internet address:
profdonpage@gmail.com}
\\
Department of Physics\\
4-183 CCIS\\
University of Alberta\\
Edmonton, Alberta T6G 2E1\\
Canada
}

\date{2012 March 8}

\maketitle
\large
\begin{abstract}
\baselineskip 25 pt

Often it is assumed that a quantum state or a phase-space distribution must be normalizable.  Here it is shown that even if it is not normalizable, one may be able to extract normalized observational probabilities from it.

\end{abstract}

\normalsize

\baselineskip 23.5 pt

\newpage

Traditionally in quantum theory one requires that quantum states be normalized,
so that they give unit expectation values for the identity operator.  A
motivation for this requirement is Born's rule.  This rule implies that for a
nondegenerate observable, an Hermitian or self-adjoint operator, which can be
written as a sum of distinct real eigenvalues each multiplied by an orthogonal
projection operator (with these projection operators forming a complete set
whose sum is the identity operator), the probability of observing a particular
eigenvalue is the expectation value of the corresponding projection operator. 
For the sum of these probabilities to be normalized to be unity, by Born's rule
the sum of the expectation values of the complete set of projection operators
must also be unity.  That is the same as the expectation value of the sum of the
complete set of projection operators, the identity operator.  Thus the
expectation value of the identity operator must be unity for Born's rule to give
normalized probabilities.

However, Born's rule does not work in a universe large enough to contain
observationally indistinguishable copies of the same observational situation
\cite{Page08,Page09a,Page09b,Page10}.  For example, if there are two copies of an observer so identical that neither can tell which one he or she is, and if each copy measures the spin of an electron, and if the quantum state is such that one copy definitely measures spin up and the other copy definitely measures spin down, the natural projection operators one would use to describe this (e.g., for either spin to be up or to be down, or for there to be at least one spin up or down) would have expectation values either 0 or 1.  However, since each observer is uncertain which copy he or she is, the probability of observing spin up should be between 0 and 1 (e.g., 1/2 if there is an equal probability to be either copy), which does not agree with Born's rule for any of the natural projection operators.  (One can rule out using the expectation values of other rather unnatural projection operators, such as for one of the spins to be horizontal, which would have expectation value 1/2, by considering other possible quantum states \cite{Page09a,Page09b,Page10}.)

Therefore, one needs some alternative to Born's rule to extract observational
probabilities from a quantum state in a sufficiently large universe.  The
simplest class of generalizations of Born's rule would seem to be that the
probabilities of observations are the normalized expectation values of as-yet
unknown positive operators (operators with eigenvalues that are real and
nonnegative) that are not necessarily projection operators.  Let us call these
operators {\it observation operators}.  

The sum of the expectation values of these observation operators should be
normalizable in order to give normalized observational probabilities.  However,
once Born's rule is abandoned, the sum of the observation operators need not be
the identity, so one no longer has the requirement that the expectation value of
the identity operator be normalizable. That is, the quantum state, interpreted
as a linear functional giving expectation values of operators, need not be
normalizable.

In more detail, if quantum operators are taken to be operators acting on an countably-infinite-dimensional Hilbert space with orthonormal basis vectors $|i\rangle$ for $i$ running from 1 to $\infty$, then each operator $O$ can be taken to have the form
\begin{equation}
O = \sum_{i=1}^\infty\sum_{j=1}^\infty O_{ij} |i\rangle \langle j|
\label{O}
\end{equation}
where the $O_{ij}$ are a set of complex numbers, the components of the operator $O$ in this basis.  The operator $O$ would be an observable if it were Hermitian or self-adjoint, $O = O^\dagger$ or $O_{ij}=O_{ji}^*$, where the dagger denotes the Hermitian conjugate (complex conjugate of the transpose) and the star denotes complex conjugation.  By a unitary transformation of the basis vectors to the orthonormal eigenvectors $|i'\rangle$ of the Hermitian operator $O$, this observable can be written as
\begin{equation}
O = \sum_{i'=1}^\infty r_{i'} |i'\rangle \langle i'|
\label{observable}
\end{equation}
with real coefficients $r_{i'}$, the eigenvalues of the observable $O$, and with $|i'\rangle \langle i'|$ the corresponding projection operator made up from the eigenvectors $|i'\rangle$ of the Hermitian operator $O$.

If the quantum state (expressed as a density matrix) is taken to be
\begin{equation}
\rho = \sum_{i=1}^\infty\sum_{j=1}^\infty \rho_{ij} |i\rangle \langle j|,
\label{state}
\end{equation}
with $\rho = \rho^\dagger$ or $\rho_{ij}=\rho_{ji}^*$, then the expectation value of a general operator $O$ is
\begin{equation}
\langle O \rangle = tr(\rho O) = \sum_{i=1}^\infty\sum_{j=1}^\infty \rho_{ij} O_{ji}.
\label{expectation}
\end{equation}
Born's rule gives the probability of getting the eigenvalue $r_{i'}$ of the Hermitian operator or observable $O$ as
\begin{equation}
P(i') = tr(\rho |i'\rangle \langle i'|) = \sum_{i=1}^\infty\sum_{j=1}^\infty \rho_{ij} \langle i'|i \rangle \langle j|i' \rangle.
\label{prob}
\end{equation}
Since
\begin{equation}
\sum_{i'=1}^\infty |i'\rangle \langle i'| = I,
\label{identity}
\end{equation}
the identity operator, the sum of these probabilities is then
\begin{equation}
\sum_{i'=1}^\infty P(i') = tr(\rho I) = tr(\rho) = \sum_{i=1}^\infty \rho_{ii},
\label{probsum}
\end{equation}
which for the normalization of probabilities must be unity in traditional quantum theory.

However, suppose Born's rule is replaced by a rule that the probability of observation $J$ (with $J$ running from 1 to $N$, say, where $N$ may or may not equal the dimension of the Hilbert space that here is taken to be infinite) is given by the normalized expectation value of the positive observation operator $O_J$ (an Hermitian or self-adjoint operator with nonnegative real eigenvalues),
\begin{equation}
P(J) = \frac{p_J}{\sum_{K=1}^N p_K},
\label{normprob}
\end{equation}
where the unnormalized probability is
\begin{equation}
p_J = \langle O_J \rangle = tr(\rho O_J).
\label{relprob}
\end{equation}
Then, with the total observation operator being $O_{\mathrm {total}} = \sum_{K=1}^N O_K$, the sum of all the observation operators, the only requirement is that
\begin{equation}
\sum_{K=1}^N p_K = \sum_{K=1}^N tr(\rho O_K) = tr(\rho O_{\mathrm {total}})
\label{requirement}
\end{equation}
be finite, which is much weaker than the requirement $tr(\rho) = 1$ that one gets when Born's rule is assumed.  One does not even require that $tr(\rho)$ be finite, so that the quantum state need not be normalizable.

Consider the example of a closed system (e.g., the entire universe) with a countable infinity of orthonormal basis vectors $|i\rangle$ (which, for concreteness, one might think of as energy eigenstates if the universe were asymptotically flat).  If there are $N = \infty$ observation operators that each have the form, as a simple toy example,
\begin{equation}
O_J = 2^{-J} |J\rangle \langle J|,
\label{obsop)}
\end{equation}
then even if the quantum state had the unnormalized maximally mixed form
\begin{equation}
\rho = I = \sum_{i=1}^\infty |i\rangle \langle i|,
\label{unnormstate)}
\end{equation}
one would still get normalized observational probabilities, in this case
\begin{equation}
P(J) = p_J = tr(\rho O_J) = 2^{-J}.
\label{normprobforunnormstate}
\end{equation}

One might think of this quantum state as corresponding to a system at infinite temperature, so that each energy eigenstate has equal quantum measure in the quantum state.  One might suppose that each energy eigenstate leads to some probability of a corresponding observation (say of what the energy is, as observed from within the system, since the system is closed and hence does not have external observers), but that this probability decreases with the energy of the eigenstate.  (One might imagine that the higher the energy, the harder it is for an observer to exist within the system.)  Therefore, even though the sum of the quantum measures for the different energy eigenstates diverges, the sum of the probabilities of the observations can remain finite (and be normalized to unity).

Of course, there are many other possible forms for the observation operators that would give normalized probabilities for observations even in the maximally mixed unnormalized quantum state $\rho = I$.  All that is required for this state is that the trace of the total observation operator, $tr(O_{\mathrm {total}})$, be finite.  For example, one could have $O_J = |J\rangle \langle J|$ for integers $J$ from 1 though $N$ that is finite rather than infinite, so that $O_{\mathrm {total}} = \sum_{J=1}^N |J\rangle \langle J|$ has finite trace $N$, and then $p_J = 1$ and $P(J) = 1/N$.  

Also, the observation operators need not be orthogonal, and indeed I would not
expect them to be.  For example, I would not expect the observation operator for
my current observation to be orthogonal to the observation operators for my past
observations.  If all the observation operators were orthogonal, then there
would exist quantum states in which only one observation occurred (with unit
probability), and all other possible observations would have zero probability of
occurring.  But it would seem implausible for my present observation, with all
its memories of an apparent past, to be able to exist within the actual
dynamical laws of physics without the existence of other real observations in
the past, both by others and by myself.  (I am not doubting the logical
possibility that I might observe an apparent memory that seems to be about your
or my existing in the past without your or my actually existing and having real
observations then, but I suspect that this logical possibility would be
inconsistent with the actual dynamical laws of physics, including the unknown
laws of what the observation operators are, even if the quantum state were
allowed to be different from what it actually is.)

An example of observation operators that are not orthogonal would be
\begin{equation}
O_J = (1/J^3) \sum_{K=1}^J \sum_{L=1}^J |K\rangle \langle L|,
\label{obsopnonorthogonal)}
\end{equation}
with components $O_{Jij} = 1/J^3$ for $1 \leq i \leq J$ and $1 \leq j \leq J$ and $O_{Jij} = 0$ otherwise.  These observation operators give unnormalized $p_J = 1/J^2$ in the maximally mixed quantum state and hence normalized observation probabilities $P(J) = 6/(\pi^2 J^2)$.

It may be instructive to consider the classical analogue of these quantum considerations.  A statistical classical analogue of the quantum state is a phase-space distribution, a nonnegative weight $w(p,q)$ over the classical phase space with the momenta and positions here symbolically denoted by $p$ and $q$.  Normally, one requires that the integral of the distribution over the entire phase space be unity, $\int w(p,q) dp dq = 1$, where $dp$ and $dq$ denote infinitesimal intervals for all the momentum and position coordinates.  Then if the phase space is divided into a countably infinite set of nonoverlapping cells, each labeled by the integer $i$, one could say that if an external observer looks to see where the system is within the phase space, the probability $P(i)$ that it is in the region $i$ would be the integral of $w(p,q) dp dq$ over that region.  It is for the sum of these probabilities $P(i)$ to be normalized to unity, $\sum_{i=1}^\infty P(i) = 1$, that one traditionally requires that the phase space distribution be normalized, $\int w(p,q) dp dq = 1$.

However, if one considers an isolated system (such as the entire universe is
usually thought to be), with all observations internal to the system, the
observations need not have probabilities that are simply the integrals of the
phase-space distribution over a corresponding region or cell of the phase
space.  It might be that different regions of phase space are inherently more or
less conducive for observers, so that the probabilities of observations differ
from the integrals of the phase-space distribution over the different regions by an observation-selection effect.  Then the classical analogue of the quantum observation operator $O_J$ would be a nonnegative real {\it observation function} $O_J(p,q)$ over the phase space, say for integers $J$ from 1 to $N$, giving the inherent probability density for the observation $J$ to occur at the phase-space location $(p,q)$ if indeed the system were at that location.  

When one folds in the phase-space distribution $w(p,q)$ that is the classical analogue of the quantum state $\rho$, one gets that
\begin{equation}
P(J) = \frac{p_J}{\sum_{K=1}^N p_K},
\label{normprob2}
\end{equation}
where the unnormalized probability in the classical case is
\begin{equation}
p_J = \int O_J(p,q) w(p,q) dp dq.
\label{relprobclass}
\end{equation}
Then, with the total observation function being $O_{\mathrm {total}}(p,q) = \sum_{K=1}^N O_K(p,q)$, the sum of all the observation functions, the only requirement is that
\begin{equation}
\sum_{K=1}^N p_K = \sum_{K=1}^N \int O_K(p,q) w(p,q) dp dq
= \int O_{\mathrm {total}}(p,q) w(p,q) dp dq
\label{normreq}
\end{equation}
be finite, which is much weaker than the requirement $\int w(p,q) dp dq = 1$ that one gets for an ideal external observer.  One does not even require that $\int w(p,q) dp dq$ be finite, so that the phase-space distribution need not be normalizable.

The classical analogue of the unnormalized maximally mixed quantum state is the phase-space distribution $w(p,q) = 1$, which is not normalizable in the usual case (assumed here) in which the phase space is infinite.  Under the traditional interpretation that the phase-space distribution gives the probabilities for ideal measurements by external observers, this uniform distribution over the phase space would not be allowed, but for observations within a closed system, it would be consistent with normalized observational probabilities so long as the total observation function is integrable, $\int O_{\mathrm {total}}(p,q) dp dq$ finite.

Just as in the quantum case with an unnormalizable maximally mixed state in which there are many ways to make the total observation operator have finite trace, so in the classical case with a uniform phase-space distribution there are also many ways to make the total observation function integrable over the entire phase space.  For example, one could take an infinite sequence of phase space regions that each have fixed phase-space volume $V = \int dp dq$, with each region labeled by $J$ that runs from 1 to $N = \infty$, and then take $O_J(p,q) = 2^{-J}/V$ in each respective region (and zero outside), which gives $\int O_{\mathrm {total}}(p,q) dp dq = 1$ and hence the normalized probabilities $P(J) = p_J = 2^{-J}$.  Alternatively, one could take a finite sequence of such phase space regions, labeled by $J$ that runs from 1 to finite $N$, and take $O_J(p,q) = 1$ inside each respective region (and zero outside), which gives $\int O_{\mathrm {total}}(p,q) dp dq = N V$ and normalized probabilities $P(J) = p_J/(NV) = 1/N$.

Just as there is no good reason I can see in the quantum case to assume that the different observation operators are orthogonal, similarly I see no good reason in the classical case to assume that the different observation functions are nonzero only in different nonoverlapping regions.  For example, they might be overlapping gaussians, such as $O_J(p,q) = \exp{[-(p-p_J)^2/p_0^2 - (q-q_J)^2/q_0^2]}$, where $(p-p_J)^2$ denotes the square of the distance in momentum space from a fiducial point labeled by $p_J$, and similarly for $(q-q_J)^2$ in position space.

One big advantage of being freed from requiring phase-space distributions to
be normalizable is that this liberates one from restricting observational
probabilities to being at a given moment of time.  When one has an observer
external to the system that makes observations at definite times, then it can
make sense to consider the probabilities of different observations at a fixed
time.  However, when one considers observations within a closed system, such as
the universe, one wants the probability of the observation itself, without
having to know what the time is.  (Of course, if the time is part of the
observation, the probability can depend on it, and one can restrict to
observations at a fixed observed time to get the conditional probabilities of
other parts of the observation, given the time part, but one would like to be
able to get the absolute probabilities of all observations, and not just
conditional probabilities when the time is known and is fixed.)

For the probabilities of all observations within an evolving closed classical system that has a particular time variable, one would expect this to depend on the phase-space distribution over all times (or at least over all times at which there are observations).  In the case that the system has a time-independent Hamiltonian, one would not expect the observation probabilities to have an explicit dependence on the time, but only a dependence on the time integral of the phase-space distribution.  That is, if the phase-space distribution is actually a function of time $t$, $w(p,q,t)$ rather than simply $w(p,q)$, instead of the unnormalized observation probabilities being $p_J = \int O_J(p,q) w(p,q) dp dq$, one would expect them to have the form
\begin{equation}
p_J = \int O_J(p,q) w(p,q,t) dp dq dt = \int O_J(p,q) W(p,q) dp dq,
\label{relprobclasswithtime}
\end{equation}
where
\begin{equation}
W(p,q) = \int w(p,q,t) dt
\label{relprobclasswithouttime}
\end{equation}
is the time integral of the time-dependent phase-space distribution $w(p,q,t)$.  This makes the assumption that the observation functions $O_J(p,q)$ do not have an explicit dependence on the time.

Now even if the time-dependent phase-space distribution $w(p,q,t)$ is integrable, $\int w(p,q,t) dp dq$ finite for each value of the time $t$, generically the time-integrated phase-space distribution $W(p,q)$ will not be integrable if the time $t$ has an infinite range.  If the configuration (position) space is bounded and if the Hamiltonian is independent of time, then even $W(p,q)$ may be infinite, but in the more common case in which the configuration space is infinite, $W(p,q)$ may be finite but not integrable.  For example, for a one-dimensional configuration space with $H = p^2/(2m)$, the phase-space distribution has the general form $w(p,q,t)=f(p,pt-mq)$ for a general function $f$ of two arguments, and if, for example, one chooses it to be $w(p,q,t) = p^{2n}\exp{[-\pi(pt-mq)^2]}$ for some positive integer $n$, one gets the time integral to be $W(p,q) = |p|^{2n-1}$, which for $n \geq 1$ is finite but not integrable.

Therefore, if one wants observation probabilities that are independent of the time variable (which generically is not directly observable from within the system), in the classical case one would like a time-independent phase-space distribution like $W(p,q)$, the time integral of the time-dependent phase-space distribution $w(p,q,t)$, but such a time-independent phase-space distribution is generally not integrable:  $\int W(p,q) dp dq$ is generally infinite if the time $t$ has an infinite range.  Therefore, it is very useful to be able to have observational probabilities like those defined above that can be finite even if the phase-space distribution is not integrable.

One runs into a similar problem in canonical quantum cosmology, in which the wavefunction is a function on the configuration space that obeys the Wheeler-DeWitt equation or something similar and has no explicit dependence on time.  For the usual case of an unbounded configuration space, the analogue of the nonnormalizability of the phase-space distribution when integrated over time to make it stationary is the nonnormalizability of the absolute square of the wavefunction when integrated over the unbounded configuration space.  This nonnormalizability occurs even for the simplest minisuperspace model in which there is the single configuration-space variable $a$, the scale size of the universe:  the integral of the absolute square of the wavefunction over the infinite range of $a$ generally diverges.  However, this need not prevent one from having normalizable observational probabilities if the observation operators are no longer restricted to form a complete set of projection operators.

If indeed the observation operators (or observation functions) permit normalized observational probabilities even for an unnormalizable maximally mixed state (or for a unnormalizable uniform phase-space distribution), this might seem to exacerbate the problem of the arrow of time, as it would seem difficult to explain our observations of order and of the apparent increase of entropy if indeed the universe actually is maximally disordered.  Presumably the observation operators could be such as to favor observations of order even if the quantum state is maximally disordered, but it seems to me rather implausible to have all of the explanation rest upon the unknown observation operators and none upon the quantum state and the dynamical laws for it that we think we partially understand \cite{Page09c}.  If the quantum state of the universe is really maximally mixed, then it would seem that our apparent partial understanding of the dynamical laws of physics would actually not at all explain our observations; all of the explanation would have to come from the observation operators that we do not yet know.

One argument for a partial explanation of the observed arrow of time has been that a maximally mixed state (or a uniform phase-space distribution) is not normalizable, and that any normalizable distribution will spread out over the available phase space so that at late times it may appear to give an arrow of time \cite{CarrollChen04,CarrollChen05,Carroll08a,Carroll08b,Carroll08c,Carroll10}.  However, if there is no requirement for a normalizable quantum state or phase-space distribution, this argument seems to lose some of its force, as one reverts to the apparent mystery of why the universe does not seem to be in a maximally mixed state.

On the other hand, even if a maximally mixed state is mathematically consistent
with normalized observational probabilities, I do not see any strong reason to
assume that the universe must be in such a state.  By Occam's razor, we would
like to find the simplest theory consistent with our observations.  (More
precisely, we would like to find theories with the highest possible posterior
probability, which by Bayes' theorem is proportional to the product of the prior
probability of the theory and the probability that the theory gives for our
observations.  Although the prior probabilities are unavoidably subjective, we
generally would assign higher prior probabilities to simpler theories.)  The
maximally mixed state is certainly a simple state, so one might well assign it a
high prior probability, but it need not be the only simple state.  Even if
another state is not quite so simple as the maximally mixed state, and so is
assigned a somewhat lower prior probability, if it gives a significantly higher
probability for our observations, it can have a higher posterior probability.  

Since it seems plausible that the maximally mixed state would give mostly highly
disordered observations, it would seem that it would give a much lower
probability for our ordered observations than a suitable highly ordered state. 
If such a state can be found that is not too complicated, and hence is not
assigned too low a prior probability, I strongly suspect that it would result in
a much higher posterior probability than the maximally mixed state.  So even
though it may be possible for the maximally mixed state to be mathematically
consistent with normalized observational probabilities, I suspect that it will
turn out to be statistically inconsistent with our observations (give a much
lower posterior probability than another theory).

Having argued that in principle one can get normalized observational probabilities from an unnormalizable quantum state or phase-space distribution, I do want to admit that this still looks like a difficult task.  For example, if eternal inflation has made our universe infinitely large, and if observations depend only on what is happening in a local region that can remain the same despite infinitely many changes elsewhere in the infinite space, then it is hard to see how to keep even the observation operators finite.  The probability that an observation occurs in some region might be thought to be roughly proportional to the expectation value of the product of some nontrivial operator in this region and the identity operators in all the other regions that do not matter for the observation in the fixed region.  But the identity operators in infinitely many other regions would seem to lead to a divergence in the trace of this infinite-product operator.

Furthermore, even if one could handle this infinite-product operator, since presumably the observation could in principle occur in any of the infinitely many regions, to get the true observation operator, one would apparently need to sum over all such infinite-product operators with the nontrivial part allowed to be in any of the infinitely many regions.  (Even if the original infinite-product operator were a projection operator, the sum would not be, which is another way of seeing that Born's rule does not work in a universe large enough for observations to occur in different regions:  the probability of the observation is not given by the expectation value of a projection operator, but by something that is at least a sum of projection operators.)  This infinite sum also seems difficult to do.

One might see a glimmer of hope from the fact that the constraint equation of quantum gravity is nonlocal, so it appears to be false to take the Hilbert space to be a product of Hilbert spaces for all of the possibly infinitely many different regions.  One might hope that this fact might in the end temper the infinities that otherwise seem to arise, but it is far beyond my ability to see how to do this, so at present I shall just admit that although it seems to be possible to have normalized observational probabilities even if the quantum state or phase-space distribution is not normalizable, I do not know in detail how to accomplish this.  Of course, I also do not know to accomplish it even if the quantum state is normalizable, so the fact that I do not know how to define normalized observational probabilities at all should not be taken as evidence against the possibility raised here that the quantum state or phase-space distribution need not be normalizable.

I am grateful for discussions with many colleagues, including Andreas Albrecht,
Thomas Banks, Raphael Bousso, Sean Carroll, Brandon Carter, Benjamin Freivogel,
Gary Gibbons, Alan Guth, James Hartle, Thomas Hertog, John Leslie, Andrei Linde,
Donald Marolf, Roger Penrose, Martin Rees, Leonard Susskind, Neil Turok, William
Unruh, Vitaly Vanchurin, Alexander Vilenkin, Robert Wald, and Edward Witten.  I
also appreciated the hospitality in India of Salaam Balaak Trust in Delhi, World
Vision in Alwar, Metropolitan Mission in Vijayawada, Missionaries of Charity in
Kolkata, and hotels in Delhi, Agra, Jaipur, Alwar, Vijayawada, and Varanasi,
where some of these ideas arose during times of touring India and visiting
various nonChristian and Christian humanitarian relief efforts there.  This work
was supported in part by the Natural Sciences and Engineering Council of Canada.

\end{document}